\documentclass[psfig]{aa}
\begin{document}

\title{Monte Carlo transition probabilities. II. }

% \subtitle{}

\author{L.B.Lucy}

\offprints{L.B.Lucy}

\institute{Astrophysics Group, Blackett Laboratory, Imperial College 
London, Prince Consort Road, London SW7 2AZ}

\date{Received ; accepted }

\maketitle

\begin{abstract}

The macroscopic quantizations of matter into macro-atoms and radiant and
thermal energies into $r$- and $k$-energy packets initiated in Paper I is
completed
with the definition of transition probabilities governing energy
flows to and from the thermal pool. The resulting Monte Carlo method is then
applied to the problem of computing the hydrogen spectrum of a Type II
supernova. This test problem is used to demonstrate the scheme's consistency
as the number of energy
packets $ {\cal N} \rightarrow \infty$, to investigate the accuracy of 
Monte Carlo estimators of radiative rates, and to illustrate the convergence
characteristics of the geometry-independent, constrained $\Lambda$-iteration
method employed to 
obtain the NLTE stratifications of temperature and level populations. In
addition, the
method's potential, when combined with analytic ionization and excitation 
formulae,  for obtaining useful approximate NLTE solutions is
emphasized.


\end{abstract}

\section{Introduction}
In an earlier paper (Lucy 2002; Paper I), a technique for imposing the
constraint of
statistical equilibrium on Monte Carlo (MC)
transfer codes was developed. This was achieved by first replacing 
the rate-equation representation of statistical equilibrium by
one involving macroscopic energy flows and then quantizing these flows into
indivisible 
energy packets. These $e$-packets - the MC quanta - 
contain either radiant
energy ($r$-packets) or kinetic energy ($k$-packets) and undergo changes on
interaction with matter in
accordance with certain probabilistic rules. Because these rules are
analogous to
the transition probabilities that govern a real atom's interactions with
photons and electrons (or other colliding species), they can be 
interpreted as (MC) transition probabilities that govern a
{\em macro-atom's} interactions with $r$- and $k$-packets,
with the macro-atom simply
representing the atoms of a particular species in a finite volume element.
     
	In addition to providing its theoretical basis,
Paper I subjected the macro-atom formalism to various numerical tests, 
each of which concerned a single point in a stratified atmosphere. In the
first group of tests, the consistency of the scheme was confirmed by
demonstrating that, when used with the exact level
populations, the technique asymptotically reproduces the correct line- and
continuum emissivities. But consistency, though necessary, does not guarantee
that an effective NLTE code can be developed with this technique. 
Accordingly, Paper I also tested the sensitivity of the MC 
emissivities to departures from the exact level populations, since 
extreme sensitivity would make it unlikely that the scheme would be
successful in an iterative search for a NLTE solution. Fortunately,
the emissivities were found to be remarkably insensitive to the level
populations, thus suggesting good convergence characteristics.

	Given these successes with one-point tests, the next step is
evidently to apply the macro-atom formalism to a stratified medium. 
Possible test problems are many and varied since
the technique is quite general: potentially, it applies to
multi-species
plasmas, to multi-dimensional geometries, and to problems with 
non-radiative heating. But given the technique's novelty, step-by-step
testing is advisable, starting with the simplest of problems.
Accordingly, this paper treats a 1-D medium comprising just one
atomic species and subject only to radiative heating.
Specifically,
the problem is to compute the level populations and kinetic temperature
throughout the outer envelope of a Type II supernova, which we take to
be pure hydrogen. In such extended, low density envelopes, substantial 
departures from LTE arise due to the strong dilution of continuum
radiation and to the negligible importance of collisional excitations. 
The technique's ability to
solve non-trivial NLTE problems is therefore tested. In particular, the
convergence behaviour of
the iterations required to obtain the level populations and the
temperature stratification can be investigated. Moreover, this test serves to
illustrate the structure of NLTE codes using the macro-atom formalism.

\section{Supernova model}   
In this section, the basic assumptions concerning the structure of the 
supernova and the relevant atomic physics are stated. In contrast to previous
MC codes for SNe (Lucy 1987; Mazzali \& Lucy 1993; Lucy 1999b; 
Mazzali 2000), the interest 
here is to test technique rather than to construct a code immediately 
suitable for 
analysing  observational data. Accordingly, state-of-the-art modelling 
precision with respect to the SN envelope or to the atomic physics and
radiative transfer is of no concern. 

\subsection{Envelope stratification}   

In formulating this test problem, we take the computational domain $D$ to be 
exterior to the layers where the energy released by radioactive decays is
thermalized. Then, with the further
assumptions that the radiative cooling and recombination time scales are
small compared to the expansion time scale, the principles governing spectrum
formation are statistical and thermal equilibrium in every local co-moving
frame within $D$. 

	Because the energy sources are interior to the lower
boundary ($r = R$) of $D$, this calculation does not predict the SN's
luminosity $L$, which is therefore a parameter. For the same
reason, a boundary condition at $r = R$ on
the outwardly-directed radiation field is required. This is taken to be   
\begin{equation}
 I^{+}_{\nu}(R) = B_{\nu}(T_{b})   
\end{equation}
where $T_{b}$, the black-body temperature, remains to be determined.

	In Eq. (1) and throughout this paper, $\nu$ denotes frequency
in the matter frame ; frequency in the rest frame will be denoted by 
$\nu_{R}$. Similarly, the energy of an $r$-packet in the matter and rest
frames will be denoted by $\epsilon_{\nu}$ and $\epsilon_{R}$,
respectively.

	The density distribution at time $t$ after the explosion is obtained
as usual from the assumption of homologous expansion. Thus
\begin{equation}
 \rho(v,t) = \left( \frac{t_{1}}{t} \right)^{3} \rho_{1}(v) 
\end{equation}
where $v = r/t$ is the constant velocity of the mass layer that is at radius
$r_{1}$ at time $t_{1}$, and $\rho_{1}(v)$ is the density-velocity profile at
elapsed
time $t_{1}$. The profile adopted is that of Arnett's (1988) explosion
model for SN1987A.

	With the further assumption of a pure H composition,
the atomic number density is $n_{H}(r,t) = \rho(r,t)/m_{H}$.

\subsection{Parameters}   

	For the problem thus defined, the basic parameters
are $t$ and $L$. The test calculation of Sect. 7 is
carried out at $t = 9.8$ days, and $L(t)$ is required to be
$1.68 \times 10^{41}$ erg s$^{-1}$, the value determined observationally
for SN 1987A (Suntzeff \& Bouchet 1990).

	In addition to setting parameters, the location of the lower boundary
must be specified. This is taken to be the layer moving at
$v_{1} = 6500$ km s$^{-1}$, corresponding to
$R = 5.50 \times 10^{14}$ cm. 

\subsection{Discretization}   

	The MC calculation is performed for a discretized version of the
SN's envelope. Specifically, the envelope is modelled as ${\cal M}$
constant-density spherical shells, within each of which the temperature
$T$ and the level populations $n_{i}$ are also constant.

	The independent variable is $x = R/r$, in terms of which
the shells have constant thickness
 $\Delta x = (x_{1}-x_{{\cal M}+1})/{\cal M}$,
with $x_{1} = 1$ and $x_{{\cal M}+1} = 0.25$. The mean radius of the $m$th
shell is
defined to be $\overline{r}_{m} = R/ \overline{x}_{m}$, where
$(\overline{x}_{m})^{-3} = (x_{m}^{-3}+x_{m+1}^{-3})/2$. The density of the
$m$th shell is then the value given by Eq. (2) at
$v=\overline{r}_{m}/t$.

\subsection{Atomic model}   

The model H atom is that used in the one-point consistency test in
Sect. 5.2 of Paper I. Thus there are 15 levels, with level 15
being the H$^{+}$ continuum $\kappa$. The 14 bound levels correspond to
principal quantum numbers $i = 1-14$ and have consolidated statistical
weights $g_{i} = 2i^{2}$.

	Transition probabilities and photoionization cross sections for
hydrogen can of course be computed with arbitrary accuracy. Accordingly,
the oscillator strengths for bound-bound (b-b) transitions are derived from 
an accurate look-up table. But the continuous absorption coefficient
is simplified by setting Gaunt factors $= 1$. Thus, the cross section
for photoionization from level $i$ is
\begin{equation}
 a_{i \kappa}(\nu) = a_{i} \left( \frac{\nu_{i}}{\nu}\right)^{3} \;\;  
 \mbox{for}  \;\; \nu > \nu_{i}
\end{equation}
where $a_{i}$ is the cross section at the ionization threshold frequency 
$\nu_{i}$.

\section{Monte Carlo preliminaries}   

In the following three sections, a detailed account is given of how the
internal radiation field in the SN's envelope is derived using the macro-atom 
formalism of Paper I. This MC calculation is subsequently embedded in an
outer iteration loop (Sect. 6.5) in the search for the NLTE solution.

\subsection{Random numbers}   
 
A random number generator is essential for every MC code.
In this investigation, the routine {\bf ran2} from Numerical
Recipes (Press et al. 1992) is used, but with minor changes to convert to
double precision. Independent tests confirm the high quality of sequences of
random numbers created with this routine.

	In the rest of this paper, the variable $z$ denotes a random
number sampling a uniform distribution in the interval $(0,1)$ and thus 
indicates a call to {\bf ran2} in the MC code.

\subsection{Formulation}   

In Paper I, three formulations of the macro-atom
technique were presented. Two concerned alternative treatments of stimulated
emission; and the third, moving media in the Sobolev limit. Here,
in view of the large velocity gradients
in a SN envelope, we treat line formation in the Sobolev
approximation and so follow the formulation presented in Sect. 4.3
of Paper I.

	Although accuracy is not of concern in this paper, it is
nevertheless of
interest that Duschinger et al. (1995) find that, in treating the
formation of hydrogen
lines in type II SNe, the Sobolev approximation is as accurate as the
comoving frame method.

\subsection{Approximations}   

	Because of the envelope's large optical depth in the Lyman
continuum, an $r$-packet in that continuum will propagate a negligible
distance before it undergoes a bound-free (b-f) absorption. Accordingly,
a useful economy of computational effort is
obtained by assuming that such packets are
re-absorbed at the point of emission. This
is the familiar `on-the-spot' approximation commonly used for models of
photoionized nebulae (e.g., Osterbrock 1974). Moreover, as for such models,
this approximation is implemented for the dominant source of
$r$-packets with $\nu > \nu_{1}$ by setting $\alpha_{1}$ and $\gamma_{1}$,
the ground state
recombination and photoionization coefficients, to zero. Of course, Lyman
continuum $r$-packets can in principle also  be emitted by f-f emission and
by recombinations
to excited levels. Such packets if they occur are re-absorbed in situ by
activating a macro-atom to state $\kappa$.  

	The above approximation effectively imposes an upper limit
$\nu_{U} = \nu_{1}$ on the MC radiation field. In a second approximation,
a lower limit $\nu_{L}$ is also imposed. The reason for this is the $\nu^{-2}$
dependence of f-f absorption as $ \nu \rightarrow 0$, which results in
occasional jumps in estimates of ${\cal H}^{ff}$, the f-f heating rate, when
an $r$-packet is
emitted with $\lambda \ga 100 \mu m$. The limit $\nu_{L}$ eliminates such
jumps at the price of a slightly biased estimate of ${\cal H}^{ff}$.

	The chosen lower limit $\nu_{L} = \nu_{U}/196$, the ionization
threshold of the highest bound level $i = 14$. To ensure that $r$-packets
are not emitted with $\nu < \nu_{L}$, we set oscillator strengths
$f_{ji} = 0$ if $\nu_{ji} < \nu_{L}$ and also exclude $\nu < \nu_{L}$ when
sampling the f-f emissivity function.

	Note that, in order to avoid loss of generality, these devices for
imposing frequency limits are ignored in all subsequent formulae.  

	A further approximation is the neglect in the transfer theory of all
terms of $O(v/c)$ except for the Doppler effect (McCrea \& Mitra 1936;
Lucy 1971,1999b). But apart from time-delay, it is quite
straightforward to include relativistic effects in MC codes
(Mazzali \& Lucy 1993).

\subsection{Initiation}   

	The NLTE solution must be obtained iteratively and so starting
values are required for the basic variables. For every
iteration except the first, the previous iteration's radiation
field $J_{\nu}$ is available, as also are the
corresponding temperatures $T$ and level-populations $n_{i}$.
For the first iteration, initial guesses are required for
these quantities.   

	The initial mean intensity of the radiation field
in the Balmer and higher continua is taken to be 
\begin{equation}
 J_{\nu}(r) = W(r) B_{\nu}(T_{b})   
\end{equation}
where the dilution factor $W = \frac{1}{2} [1-\sqrt{1-x^{2}}]$.

	The lower-boundary temperature $T_{b}$, which also determines the
frequency
distribution of $r$-packets launched into $D$ at $r = R$ - see Eq. (1), is
initiated at $T_{b} = 6000$K.

	With the radiation field specified, values of $T$,
$n_{i}$ and $n_{e}$ in each shell
could be obtained by
imposing the constraints of statistical and thermal equilibrium as
described in Sect. 6. Instead, the simpler and cruder option is followed of
assuming an
isothermal stratification with $T =5000$K and then obtaining the
initial $n_{i}$ and $n_{e}$ from the Saha and Boltzmann
formulae.

\subsection{Monte Carlo transition probabilities}   

	With the current estimates of $J_{\nu}$, $T$ and $n_{i}$, 
the MC transition probabilities are computed for the
next simulation of the radiation field. Following an $e$-packet's 
capture by a macro-atom, these probabilities determine stochastically the
characteristics of the subsequently-emitted $e$-packet.

	The MC transition probabilities defined in Paper I and in this
paper involve {\em ratios} of the rates of various radiative and collisional
processes. As in Paper I, the formulae for such rates always refer to unit
volume.

	Adopting the notation of Paper I, we define 
${\cal R}_{ij} = R_{ij}+ C_{ij}$ to be the total rate of the
transition $i \rightarrow j$, with $R_{ij}$ and $C_{ij}$ denoting 
radiative and collisional rates. Also, except where noted,
the summation convention of Paper I is assumed. Thus suffixes $\ell$ and $u$
indicate summations over all levels $<i$ and all levels $>i$,
respectively.    

The probability that a
macro-atom in state $i$ emits an $r$-packet and thus de-activates is then 
\begin{equation}
 p_{i}^{r}= R_{i \ell} (\epsilon_{i} - \epsilon_{\ell}) / {\cal D}_{i}
 \;\;\;\; \mbox{with} \;\;\;\;
                 {\cal D}_{i} = ({\cal R}_{i\ell}+{\cal R}_{iu})\epsilon_{i}
\end{equation}
where $\epsilon_{i}$ is the excitation energy of
level $i$. 
The corresponding probability for the emission of a $k$-packet is 
\begin{equation}
 p_{i}^{k}= C_{i \ell} (\epsilon_{i} -  \epsilon_{\ell})/ {\cal D}_{i} 
\end{equation}

	In addition to the options of emitting an $r$- or a $k$-packet,
a macro-atom in state $i$ can spontaneously jump to any other state
$j$. Moreover, such internal transitions occur without the absorption or
emission of an $e$-packet. The probabilities of internal transitions to 
the {\em particular} upper or lower states $u$ and $\ell$ are
\begin{equation}
 p_{iu}={\cal R}_{iu}\epsilon_{i}/{\cal D}_{i} \;\;\;\;\; \mbox{and}
\;\;\;\;\; p_{i \ell}= {\cal R}_{i \ell}\epsilon_{\ell}/{\cal D}_{i} \;\;\; 
\end{equation}

	The above options for leaving state $i$ are complete; consequently,
\begin{equation}
 p^{r}_{i} + p^{k}_{i} + p_{iu} +  p_{i \ell} = 1
\end{equation}
as may readily be demonstrated.

	In order to compute these MC transition probabilities, the rates
$R_{ij}$ and $C_{ij}$ must be specified.

\subsubsection{Collisional rates}   

The rates of excitation and de-excitation due to collisions
by thermal electrons are
\begin{equation}
 C_{ji} = q_{ji}n_{j}n_{e}  \;\; \mbox{and} \;\;  C_{ij} = q_{ij}n_{i}n_{e}
\end{equation}
The rates of collisional ionization and recombination have the same form,
except that the latter, being a three-body process, is $\propto n_{e}^{2}$.

	For excitations, the coefficient $q_{ji}$
is computed with van Regemorter's (1962) approximate formula and the
inverse coefficient $q_{ij}$ by detailed balancing. For collisional
ionizations, the coefficient $q_{j \kappa}$ is evaluated with the
approximate formula given by Mihalas (1978, eq. [5-79]), with the inverse
coefficient $q_{\kappa j}$ again obtained by detailed balancing.

\subsubsection{Radiative rates}   

The radiative bound-bound (b-b) rates are computed
in the Sobolev approximation as discussed earlier. Thus, with stimulated
emissions treated as negative absorptions,
the radiative rates
between bound level $i$ and a lower level $j$ are 
\begin{equation}
 R_{ij} = A_{ij}\beta_{ji} n_{i}  \;\;\; \mbox{and}
\;\;\;  R_{ji} = (B_{ji}n_{j}-B_{ij}n_{i})\beta_{ji} J_{ji}^{b} 
\end{equation}
Here the mean intensity $J_{ji}^{b}$ refers to the far blue wing  of the
transition
$j \rightarrow i$, and $\beta_{ji}$ is the Sobolev escape probability,
given by
\begin{equation}
 \beta_{ji}= \frac{1}{\tau_{ji}}(1-e^{-\tau_{ji}}) 
\end{equation}
where $\tau_{ji}$ is the transition's Sobolev optical depth. In the
special case of homologous expansion,
\begin{equation}
 \tau_{ji}= (B_{ji}n_{j}-B_{ij}n_{i})\frac{hct}{4\pi} 
\end{equation}
with $t$ being the elapsed time since the SN exploded.

	For the first iteration, the mean intensities $J_{ji}^{b}$ 
are from Eq. (4). For subsequent iterations, they are evaluated from
the MC radiation field - see Sect. 6.2.  

	The radiative processes coupling bound levels to the continuum are
photoionizations and radiative recombinations. The rate coefficient for
spontaneous recombinations to level $i$ is (e.g., Mihalas 1978, p.131) 
\begin{equation}
 \alpha^{sp}_{i} = 4 \pi \:  \Phi_{i \kappa}   
 \int_{\nu_{i}}^{\infty}
 \frac{a_{i \kappa}(\nu)}{h \nu} \;
 \frac{2h \nu^{3}}{c^{2}} \;
             e^{-h \nu/kT} \; d \nu
\end{equation}
with 
\begin{equation}
 \Phi_{i \kappa}(T)= \frac{n_{i}^{*}}{n_{\kappa}^{*} n_{e}}   
\end{equation}
where an asterisk indicates the level's LTE population at the given $T$
and $n_{e}$. The corresponding rate coefficient for stimulated recombinations
$\kappa \rightarrow i$ is
\begin{equation}
 \alpha^{st}_{i} = 4 \pi \:  \Phi_{i \kappa}   
 \int_{\nu_{i}}^{\infty}
 \frac{a_{i \kappa}(\nu)}{h \nu} \: J_{\nu} \: 
             e^{-h \nu/kT} \: d \nu
\end{equation}
and that for photoionizations $i \rightarrow \kappa $ is
\begin{equation}
  \gamma_{i}
 = 4 \pi \int_{\nu_{i}}^{\infty}
 \frac{a_{i \kappa}(\nu)}{h \nu}  J_{\nu} d \nu
\end{equation}
In terms of the above, the total recombination coefficient for level $i$ is
$\alpha_{i} = \alpha_{i}^{sp}+\alpha_{i}^{st}$.

	If we treat stimulated recombinations as negative photoionizations,
the radiative rates coupling $i$ and $\kappa$ are
$R_{\kappa i}= \alpha_{i}^{sp} n_{\kappa} n_{e}$ and
$R_{i \kappa}= \gamma_{i} n_{i}- \alpha_{i}^{st} n_{\kappa} n_{e}$. This
latter rate may be written as $R_{i \kappa}= \tilde{\gamma}_{i} n_{i}$, where
the corrected photionization coefficient
\begin{equation}
  \tilde{\gamma}_{i}
 = 4 \pi \int_{\nu_{i}}^{\infty}
 \frac{\tilde{a}_{i \kappa}(\nu)}{h \nu}  J_{\nu} d \nu
\end{equation}
is expressed in terms an absorption cross section 
\begin{equation}
 \tilde{a}_{i \kappa}(\nu) = a_{i \kappa}(\nu) 
 (1 -  \frac{n_{\kappa}}{n_{i}}
           \frac{n_{i}^{*}}{n_{\kappa}^{*}} e^{-h \nu/kT}) 
\end{equation}
which incorporates the NLTE correction factor for stimulated emissions.

\subsection{Modified rate coefficients}   

The rates at which b-f and f-b transitions absorb and emit
radiant energy are needed in the course of the calculation. These can
conveniently be expressed in terms of modified photoionization and
recombination coefficients. Thus the rate at which
photoionizations from level $i$ remove energy from the radiation field 
is $ \gamma_{i}^{E} \: h \nu_{i} \: n_{i} $, where  	
\begin{equation}
  \gamma_{i}^{E}
 = 4 \pi \int_{\nu_{i}}^{\infty}
 \frac{a_{i \kappa}(\nu)}{h \nu_{i}}  J_{\nu} d \nu
\end{equation}
Comparison with Eq. (16) shows that this modified coefficient
is obtained from the conventional one by changing $h \nu$ to
$h \nu_{i}$ in the integrand's denominator.

	Similarly, the corresponding rate at which spontaneous
recombinations to level $i$ add energy to the radiation field is 
$ \alpha_{i}^{E ,sp} \: h \nu_{i} \: n_{\kappa} n_{e} $, where
\begin{equation}
 \alpha_{i}^{E ,sp} = 4 \pi  \: \Phi_{i \kappa}       \:  
 \int_{\nu_{i}}^{\infty}
 \frac{a_{i \kappa}(\nu)}{h \nu_{i}} \;
 \frac{2h \nu^{3}}{c^{2}} \;
             e^{-h \nu/kT} \; d \nu
\end{equation}
This modified coefficient 
is again obtained from the conventional one [Eq. (13)]
by changing $h \nu$ by $h \nu_{i}$ in the integrand's denominator.

	Following the same procedure, the modified coefficients 
$\alpha_{i}^{E, st}$ and $\tilde{\gamma}_{i}^{E}$ are also defined.
Thus stimulated recombinations to level $i$ emit radiant energy at
the rate $ \alpha_{i}^{E, st} \: h \nu_{i} \: n_{\kappa} n_{e} $.
But when treated as negative photoionizations, these 
stimulated recombinations reduce the rate at which photoionizations
from level $i$ absorb radiant energy to
$ \tilde{\gamma}_{i}^{E} \: h \nu_{i} \: n_{i} $.

	In terms of the above, the total rate at which recombinations to
level $i$ emit radiant energy is
$\alpha_{i}^{E} \: h \nu_{i} \: n_{\kappa} n_{e} $,
where $\alpha_{i}^{E} =  \alpha_{i}^{E ,sp} + \alpha_{i}^{E ,st}$.

	For the first iteration, the coefficients $\gamma_{i}$,
$\gamma_{i}^{E}$,$\alpha_{i}^{st}$ and $\alpha_{i}^{E ,st}$ are derived by
numerical integration with $J_{\nu}$ from Eq.(4). For subsequent
iterations, they are evaluated using the MC radiation field - see Sect. 6.2.
In terms of these coefficients, the corrected photionization
coefficient 
$\tilde{\gamma}_{i}= \gamma_{i} - \alpha_{i}^{st} n_{\kappa} n_{e}/n_{i}$
and correspondingly for $\tilde{\gamma}^{E}_{i}$

\section{Emission and absorption of energy packets}   

	This MC technique directly models the flow of energy through the 
domain $D$ by computing
the interaction histories of numerous indivisible $e$-packets. In
consequence of these interactions, an $e$-packet is sometimes an   
$r$-packet and at other times a $k$-packet. Accordingly, in this section, 
the probabilistic rules governing the emission and absorption
of both $r$- and $k$-packets are developed.

\subsection{Initial creation of $r$-packets}   

	In the absence of non-radiative heating in 
$D$, the desired solution
satisfies the constraint of radiative equilibrium in the co-moving frame at
every point in $D$. This constraint is rigorously incorporated into the
macro-atom formalism
by not allowing active macro-atoms to appear or disappear spontaneously
within $D$.
This then implies that every $e$-packet's interaction history both
begins and ends as an $r$-packet crossing a boundary of $D$ 
(Sect. 7.1, Paper I). Accordingly, in the context of a
spherically-symmetrical SN envelope, the creation
of $r$-packets corresponds to a sampling of $I^{+}_{\nu}(R)$ given
by Eq. (1).

	In the previous SN and stellar wind codes, the initial
energies $\epsilon_{0}$ of the packets were equal, independent of frequency.
But now, in order
that the weak radiation field between L$\alpha$ and the Lyman limit is well
represented, $r$-packets are created in equal numbers in equal intervals of
$\ell n \: \nu$, with energies assigned in accordance with 
black body emission at $T_{b}$. 

	Despite $r$-packets no longer having equal initial energies, MC
estimators for integrals of the radiation field can be
cast in familiar form if we take $\epsilon_{0}$ to be the {\em unit}
adopted for measuring the energies of the packets.
A convenient choice is defined by the equation
\begin{equation}
 {\cal N} \frac{\epsilon_{0}}{\Delta t} = 4 \pi R^{2}  \times \sigma T_{b}^{4}   
\end{equation}
where $\Delta t$ is the 
interval simulated by the MC experiment and ${\cal N}$ is the
number of $r$-packets
launched across the lower boundary $r = R$ in time $\Delta t$.  
   
	Now consider an $r$-packet representing emission by the lower
boundary in the interval
$\Delta \ell n \nu$ at $\nu$ during time $\Delta t$. Since the flux
transported by $I^{+}_{\nu}(R)$ in this interval is
$ \pi \nu B_{\nu}(T_{b}) \Delta \ell n \nu$, this packet's energy in the
co-moving frame is given by 
\begin{equation}
  \frac{\epsilon_{\nu}}{\epsilon_{0}} = {\cal N} \: 
  \frac{\nu B_{\nu}(T_{b}) \Delta \ell n \nu  }{\sigma T_{b}^{4}/\pi} 
\end{equation}

	A total of $\cal{N}$ packets with fractional energies given by
Eq. (22) are created
{\em uniformly} in $\ell n \nu$ in the frequency interval $(\nu_{L},\nu_{U})$
- see Sect. 3.3.

	Notice that the values of $\ell n \nu$ are {\em not} obtained by
random sampling. This is the one point in the calculation where 
stratified sampling can readily and safely be used to limit MC
noise.   

	The $r$-packets must also have their initial direction cosines 
$\mu$ specified. Assuming zero limb-darkening for emission at the lower
boundary - see Eq. (1), we randomly sample this limb-darkening law by
taking $\mu = \sqrt{z}$. The packet's initial rest energy is then
$\epsilon_{R} = \epsilon_{\nu}/(1- \mu v_{1}/c)$.

\subsection{Emission of $r$-packets}   

	A macro-atom can be activated by a $k$-packet or an
$r$-packet. The MC transition probabilities
defined in Sect. 3.5 are then applied until the macro-atom de-activates.
This occurs 
with the emission of a $k$-packet or an $r$-packet. If an $r$-packet is
emitted, the next step is to assign $\nu$.
The relevant procedure depends on whether  
de-activation occurred from the continuum state or from a discrete state.

\subsubsection{De-activation from a discrete state $1 < i < \kappa$}   

	In this case, the emitted $r$-packet belongs to an emission line
arising from atomic level $i$. One of these lines must therefore be selected
randomly, with
due weight given to the lines' emissivities.
Since the total emissivity of the lines $i \rightarrow \ell$ is
$\dot{E}_{i}^{R} = R_{i \ell} (\epsilon_{i}-\epsilon_{\ell})$, where
the $R_{i \ell}$ are from Eq. (10),
the selection probability for the line $i \rightarrow j$ is 
$R_{ij} (\epsilon_{i}-\epsilon_{j})/\dot{E}_{i}^{R}$.

	If the line thus randomly selected is $i \rightarrow j$, the $\nu$
of the emitted $r$-packet is simply the transition's
rest frequency $\nu_{ji} = (\epsilon_{i}-\epsilon_{j})/h$.
Note that because we are using the Sobolev approximation in
the narrow-line limit, no sampling of an emission profile is required.

\subsubsection{De-activation from the continuum state $\kappa$}   

	In this case,
the emitted $r$-packet's $\nu$ is obtained by sampling
the energy distributions of the spontaneous recombination continua
(Sect. 4.2.3, Paper I). This can be done by first selecting a continuum and
then selecting a $\nu$ in that continuum.

	The emissivity due to spontaneous recombinations to level $i$ is
[cf. Eq. (13)]
\begin{equation}
 4 \pi j_{i}(\nu) = 4 \pi \:   \Phi_{i \kappa} \:
 a_{i \kappa}(\nu) \; \frac{2h \nu^{3}}{c^{2}} \; e^{-h \nu/kT} 
 \:n_{\kappa}n_{e} 
\end{equation}
Substituting for $a_{i \kappa}(\nu)$ from Eq. (3) and integrating from
$\nu_{i}$ to $\infty$, we find that the integrated emissivity of the
$\kappa \rightarrow i$ continuum is
\begin{equation}
 \dot{E}_{\kappa i} = 4 \pi      \: \Phi_{i \kappa} \:
 a_{i} \; \frac{2h \nu_{i}^{3}}{c^{2}} \; 
 \frac{kT}{h} e^{-h \nu_{i}/kT}  \:n_{\kappa}n_{e}
\end{equation}
The selection probability for emission in
the $\kappa \rightarrow i$ continuum is therefore 
$\dot{E}_{\kappa i}/\sum_{i}\dot{E}_{\kappa i}$. 

	If the selected continuum is $\kappa \rightarrow i$, the
next step is to sample its energy distribution. The
emitted $r$-packet's $\nu$ is thus the solution of the equation 
\begin{equation}
 \int_{\nu}^{\infty} 4 \pi j_{i}(\nu) \: d \nu  = z \: \dot{E}_{\kappa i}
\end{equation}
Substitution from Eqs. (23) and (24) reduces this equation to the simple
formula 
\begin{equation}
  \nu= \nu_{i} \: \left(1 - \frac{kT}{h \nu_{i}} \:\ell n\: z \right)
\end{equation}

	Note that $r$-packets emitted by macro-atoms de-activating from
state $\kappa$ represent only the loss of {\em ionization} energy -
see Sect.5.2 and Fig.4 in Paper I. But recombination emission also results in
the loss of
{\em thermal} energy. This loss is represented by (some of) the
conversions of $k$-packets into $r$-packets - see Sect. 4.3.2 below.

\subsection{Creation and elimination of $k$-packets}   

The MC transition probabilities defined in Paper I model the effect of atomic
levels on the frequency redistribution of absorbed radiant energy. 
Another set of probabilities must now be defined to model the corresponding
effect of interactions with the thermal pool. In this formalism,
such interactions are represented by the emission and absorption of
$k$-packets. 

	Note that in developing probabilistic rules for $k$-packets, we 
assume that
b-f and f-f absorptions as well as collisional de-excitations and
recombinations are all followed by an instantaneous re-adjustment of 
the thermal motions to a Maxwell velocity distribution. Accordingly,
when considering the elimination of $k$-packets, we use rate coefficients
that assume a Maxwell distribution for the thermal electons.    

\subsubsection{Creation of $k$-packets}   

	The interactions creating $k$-packets correspond to the
mechanisms heating the gas. 

	Collisional de-excitations and recombinations convert excitation and
ionization energy into heat. These processes are represented in this 
formalism as ${\cal A}^{*} \rightarrow k$ - i.e., by the
emission of a $k$-packet by an active macro-atom - see Eq.(6). 

	A second heating mechanism is the conversion of radiant energy into
heat by f-f absorptions. In this formalism, this corresponds to 
$r \rightarrow k$ - i.e., to the direct conversion of an $r$-packet into
a $k$-packet without the mediation of a macro-atom.

	A third heating mechanism is the conversion of radiant energy into
heat by b-f absorptions. For this mechanism, an extended discussion is
necessary since not all absorbed energy appears as heat. 

	In terms of the modified photoionization coefficient defined in
Sect. 3.6,  the rate at which b-f transitions from level $i$
absorb radiant energy is $\tilde{\gamma}_{i}^{E} h \nu_{i} n_{i}$.
But each photoionization raises the ionization energy by 
$\epsilon_{\kappa}-\epsilon_{i} = h \nu_{i}$. Accordingly, from the  
total absorbed rate, the amount $\tilde{\gamma_{i}} h \nu_{i} n_{i}$
is used in ionizing the atoms, with only the remainder
$(\tilde{\gamma}_{i}^{E}-\tilde{\gamma_{i}}) h \nu_{i} n_{i}$
available to heat the gas.

	The existence of these two channels might
suggest that an $r$-packet undergoing a b-f absorption should be split.
But this would lose the advantages of working with indivisible packets
(Paper I, Sect. 1). Instead, therefore, the absorbed energy is assigned 
randomly to one or other of the channels. Thus, with probability
\begin{equation}
 \pi_{i}^{\kappa} =  \tilde{\gamma}_{i}/ \tilde{\gamma}_{i}^{E}
\end{equation}
the $r$-packet activates a macro-atom to its continuum state $\kappa$ -
i.e., $r \rightarrow {\cal A}^{*}_{\kappa}$.
Alternatively, with probability $1 - \pi_{i}^{\kappa}$, 
the $r$-packet converts directly into a $k$-packet - i.e., $r \rightarrow k$ .
In the first case, all of
the $r$-packet's energy is converted into ionization energy; in the
second, all is contributed to the thermal pool.

	Summing over all bound states, we find that the probability that a
b-f absorption of an $r$-packet with frequency $\nu$ results in the creation
of a $k$-packet is
\begin{equation}
 \pi^{k}(\nu) =  1-\sum_{i} \; p^{bf}_{i}(\nu) \:\pi_{i}^{\kappa} 
\end{equation}
where $p^{bf}_{i}$ is the probability that the b-f absorption occurred from
level $i$. Now, since the macroscopic b-f absorption coefficient
$k^{bf}_{\nu}$ is given by
\begin{equation}
 k^{bf}_{\nu} \rho = \sum_{\nu_{i} < \nu} n_{i} \: \tilde{a}_{i \kappa}(\nu) 
\end{equation}
where the corrected cross section $\tilde{a}_{i \kappa}$ is defined by
Eq. (18), the required probability $p^{bf}_{i}$ is evidently
\begin{equation}
 p^{bf}_{i}(\nu) =  n_{i} \: \tilde{a}_{i \kappa}(\nu) \; / \: 
                                                        k^{bf}_{\nu} \rho
\end{equation}

	The probabilities defined in this subsection are $\geq 0$ if
$ \tilde{a}_{i \kappa}(\nu) \geq 0$ for all $\nu$. From Eq. (18), we see
that this inequality is violated as $\nu \rightarrow 0$ if 
$b_{\kappa}/b_{i} > 1$, where $b_{i} = n_{i}/n_{i}^{*}$ is the departure
coefficient for level $i$. In the numerical tests of Sect. 7, minor violations
of the condition $b_{i}/b_{\kappa} \geq 1$ occur for high levels of the model
H atom, and these are probably due to the neglect of the recombinations into
levels $i > 14$ and therefore of the subsequent cascade contributions to the
included levels. To avoid negative probabilities when this happens, we
set $b_{\kappa}/b_{i} = 1$ in Eq. (18) if  $b_{\kappa}/b_{i} >  1$.

\subsubsection{Elimination of $k$-packets}   

On the assumption that advection and conduction of thermal energy are
negligible compared with radiative transport (but note the possibilities of
generalization), a $k$-packet is converted {\em in situ} back into
an $r$-packet, either directly ($k \rightarrow r$) or indirectly
($k \rightarrow {\cal A}^{*} \rightarrow \cdots \rightarrow r $). The
interactions eliminating $k$-packets correspond to the mechanisms cooling
the gas, and their selection probabilities are determined by the mechanisms'
cooling rates. 

	Collisional excitations and ionizations convert heat
into excitation and ionization energy. The total cooling rate provided by
this mechanism is
\begin{equation}
 {\cal C}^{c \ell} = \sum_{i = 2}^{\kappa} \: {\cal C}^{c \ell}_{i}
 \;\;\;\; \mbox{with} \;\;\;\;
 {\cal C}^{c \ell}_{i} =  \sum_{j = 1}^{i-1} C_{ji} \:
 (\epsilon_{i}-\epsilon_{j}) 
\end{equation}

	A second cooling mechanism is the conversion of heat into radiant
energy by f-f emission. The cooling rate in this case  
(e.g, Osterbrock 1974, p.44) is 
\begin{equation}
 {\cal C}^{ff} = C\; T^{1/2}\; n_{\kappa}n_{e} 
\end{equation}
With the mean Gaunt factor set $=1$, the coefficient
$C = 1.426 \times 10^{-27}$ (cgs).

	A third cooling mechanism is the conversion of heat into radiant
energy by f-b emission. The rate at which spontaneous recombinations 
$\kappa \rightarrow i$
convert thermal {\em and} ionization energy into radiant energy is
$\alpha^{E, sp}_{i}\: h \nu_{i}\: n_{\kappa} n_{e}$ - see Sect. 3.6. But
each such recombination releases $h \nu_{i}$  of ionization energy, which is 
therefore being radiated at the rate
$\alpha^{sp}_{i}\; h \nu_{i}\: n_{\kappa} n_{e}$. Subtracting this from the 
previous expression and summing over bound states, we find that the rate 
at which heat is radiated by spontaneous f-b
transitions is
\begin{equation}
 {\cal C}^{fb, sp} = n_{\kappa}n_{e} \sum_{i} 
 (\alpha^{E, sp}_{i}-\alpha^{sp}_{i})   \; h \nu_{i}
\end{equation}

	With cooling rates thus specified, the selection probabilities are
as follows: a $k$-packet is removed by a f-b transition ($k \rightarrow r$)
with probability
\begin{equation}
 \pi^{fb} = {\cal C}^{fb, sp}/ ({\cal C}^{fb, sp} + {\cal C}^{ff}
 + {\cal C}^{c \ell})
\end{equation}
by a f-f process  ($k \rightarrow r$) with probability
\begin{equation}
 \pi^{ff} = {\cal C}^{ff}/ ({\cal C}^{fb, sp} + {\cal C}^{ff} +
 {\cal C}^{c \ell})
\end{equation}
and by a collisional process with probability $1-\pi^{fb}-\pi^{ff}$. In this
last case, the $k$-packet activates a macro-atom
($k \rightarrow {\cal A}^{*}$) and so is not directly
converted into an $r$-packet.  

	Note that the quantity subtracted in Eq. (33), namely the rate
of loss of ionization energy, is represented in this MC procedure by
macro-atoms de-activating from state $\kappa$ (Sect. 4.2.2). Also
cooling by stimulated recombinations 
does not appear in the above formulae because these recombinations, treated
as negative photoionization, are allowed for when creating $k$-packets
(Sect. 4.3.1).

\section{Monte Carlo calculation}   

The MC model of the radiation field in the SN envelope is built up from the
trajectories of $\cal{N}$ individual $r$-packets,
starting from their launch into $D$ at $r=R$ and ending with their
exit from $D$, either by escaping to
$\infty$ or by re-crossing the lower boundary.

	As an $r$-packet propagates in $D$, it will typically undergo
numerous geometrical and physical `events'. The geometrical events
are simply crossings of the boundaries of the ${\cal M}$
constant-density shells.
But the more important {\em physical} events constitute the MC technique's
treatment of the interactions of radiation and matter. 

	In describing how the packets' trajectories are computed, it suffices
to explain how to find the next event along one packet's trajectory. Thus,
we now
suppose that an $r$-packet has just undergone event $n$ in its
propagation history and that this occurred in shell $m$. The task now
is first to find the location and nature
of event $n+1$ and then to compute the frequency $\nu$ and direction of
propagation of
the $r$-packet that emerges from this event. 

\subsection{Absorption and scattering coefficients}   

	Immediately following event $n$, the macroscopic coefficients
of continuum absorption
($k_{\nu}^{bf}$, $k_{\nu}^{ff}$) and scattering ($\sigma$)
must be computed since the $r$-packet is either in a new shell or has
changed its $\nu$.

	The b-f coefficient, corrected for stimulated
emission, is given by Eq. (18), where
the $n_{i}$ are the current estimates of the NLTE level populations in
shell $m$. The f-f coefficient
$k_{\nu}^{ff}$ is calculated with the standard formula but with mean Gaunt
factor $= 1$, in accordance with the corresponding simplification for
Eq. (3). 

\subsection{Pathlength}   

If $s$ is distance from event $n$ measured along the packet's 
trajectory 
and $\tau(s)$ is the corresponding optical depth, then
the pathlength $\Delta s$ between events $n$ 
and $n+1$ is determined by the equation
\begin{equation}
 \tau(\Delta s) = \Delta \tau
\end{equation}
where the beam's exponential decay with optical depth is randomly sampled
by taking $\Delta \tau = - \ell n (z)$.

	Continuum and line processes both contribute to $\tau(s)$. The
continuum contibution is approximately
\begin{equation}
 \tau_{\nu}(s) = \chi_{\nu} \rho \: s   \;\;\;\;\; \mbox{with} \;\;\;\;\; 
                       \chi_{\nu} = k_{\nu}^{bf} + k_{\nu}^{ff} + \sigma
\end{equation}
where $\nu$ is the packet's frequency immediately following 
event $n$. Notice that the small changes in the extinction coefficient seen
by the packet as its $\nu$ decreases along the trajectory between
events $n$ and $n+1$ are neglected. This is justified since these changes
$ \rightarrow 0$ as ${\cal M}  \rightarrow \infty$.

	The contribution of lines to $\tau(s)$ is computed with the Sobolev
approximation in the narrow-line limit, according to which a line with
Sobolev optical depth $\tau_{k}$ attenuates an incident
beam by the factor $e^{-\tau_{k}}$ at the {\em point} where resonance
occurs. 
Thus $\tau(s)$ undergoes {\em discontinuous} increments  
$\tau_{k}$ at the points $s_{k}$ where the packet's  
frequency $\nu(s)$ equals one of the rest frequencies $\nu_{k}$ of the b-b
transitions. Accordingly,
\begin{equation}
 \tau(s) =  \tau_{\nu}(s) + \sum_{k} \tau_{k}  
\end{equation}
where the summation extends over all $s_{k}$ in the 
interval $(0,s)$. Moreover, at each $s_{k}$,
\begin{equation}
 \tau(s_{k}^{+}) =  \tau(s_{k}^{-}) + \tau_{k}  
\end{equation}

	The above prescription for locating event $n+1$ 
applies only if it occurs within shell $m$. If Eq. (36)
is not satisfied in shell $m$, event $n+1$ is geometrical, and 
the pathlength $\Delta s$ is then simply the distance from event $n$ to the
point where the trajectory crosses into a neighbouring shell.

\subsection{Selecting the physical event}

	If event $n+1$ occurs within shell $m$ then it is
physical, due to either continuum extinction or a line absorption.
The former is indicated if $\Delta s \neq s_{k}$; the latter otherwise.

	If the event is a continuum extinction, we decide between 
the three contributing mechanisms
using a single random number: the event is an
electron scattering if $z < \sigma/\chi_{\nu}$; failing this, the event is a
b-f absorption if $z < (\sigma + k_{\nu}^{bf})/\chi_{\nu}$; failing this, the
event is a f-f absorption.       

	Event $n+1$ is absorption in line $k$ if $\Delta s = s_{k}$.
But since $\tau (s)$ has a discontinuity at $s_{k}$, this solution
of Eq. (36) is actually recognized by finding that
$\tau(s_{k}^{-}) < \Delta \tau < \tau(s_{k}^{+})$.

\subsection{New frequency}

	With the nature of event $n+1$ determined, the next task is 
to assign a frequency to the emitted $r$-packet.

	If event $n+1$ is geometrical, the $r$-packet is simply 
continuing its existing flight path into the new shell, and so the $\nu$ at
the beginning of the `new' trajectory is identical to the $\nu$ at the end of
the `previous' trajectory, namely the value at the boundary crossing.  

	On the other hand, if event $n+1$ is physical, the nature of the
event determines the procedure to be followed in assigning $\nu$.

\subsubsection{Electron scattering}   

With electron scatterings treated as coherent in the comoving frame, 
the scattered $r$-packet's $\nu$ is identical to that of the incident packet.

\subsubsection{Bound-bound absorption}   

If event $n+1$ is absorption by the transition $\ell \rightarrow u$, 
the immediate result is a macro-atom in state $u$ - i.e.
$r \rightarrow {\cal A}^{*}_{u}$. The next step therefore is
the repeated application of
the transition probabilities of Sect. 3.5 until the macro-atom
de-activates. This was discussed at length in Paper I and
specifically
illustrated in Sect. 5.5 of that paper with the 15-level H atom used here.

	The macro-atom de-activates with the emission of either a $k$-
packet or an $r$-packet. In the former case, the $k$-packet is eliminated 
in situ as described in Sect. 4.3.2. 

	If the macro-atom de-activates with the emission of an $r$-packet,
selection of $\nu$ depends on whether it was emitted from the continuum state
$\kappa$ or from one of the bound states $1 < i < \kappa$. These cases were
both discussed in Sect. 4.2, where the selection procedures followed in
assigning $\nu$ are given.

\subsubsection{Free-free absorption}   

If event $n+1$ was a f-f absorption, the immediate result is a $k$-packet
which then in situ either converts into an $r$-packet or excites
a macro-atom.
The conversion $k \rightarrow r$ occurs by f-b emission with probability
$\pi^{fb}$ or by f-f emission
with probability $\pi^{ff}$ (Sect. 4.3.2). For f-b emission, the selection
of $\nu$ for the emergent $r$-packet is the two step procedure described in
Sec. 4.2.2.

	For f-f emission, $\nu$ is selected by sampling this mechanism's 
emissivity function - i.e., by solving the equation
\begin{equation}
 \int_{\nu}^{\infty} j_{\nu}^{ff} d \nu =
                        z \: \int_{0}^{\infty} j_{\nu}^{ff} d \nu  
\end{equation}
If the $\nu$-dependence of the velocity-averaged Gaunt factor is
neglected, $j_{\nu}^{ff} \propto e^{-h \nu/kT}$ and the solution is 
\begin{equation}
 \nu = -\frac{kT}{h} \: \ell n \: z
\end{equation}

	Besides these two direct $k \rightarrow r$ conversions, there is 
alternatively the indirect conversion
$k \rightarrow {\cal A}^{*} \rightarrow \cdots \rightarrow r$, which occurs
with probability $1-\pi^{fb}-\pi^{ff}$ and is due to collisional cooling.
In this case, the macro-atom is activated to state $i$ with probability
$ {\cal C}_{i}^{c \ell}/ {\cal C}^{c \ell}$ - see Eq. (31). Application of
the MC transition 
probabilities as in Sect. 5.4.2 then leads eventually to the emission
of an $r$-packet.

\subsubsection{Bound-free absorption}   

If event $n+1$ is a b-f absorption, the immediate result is either a
$k$-packet (Sect. 4.3.1) or a 
macro-atom in state $\kappa$. The selection probability 
is $\pi^{k}(\nu)$ for the former and $1-\pi^{k}(\nu)$ for the latter, where
$\nu$ is the frequency of the incident $r$-packet.

	If this selection results in a macro-atom in state $\kappa$, the
MC transition rules are applied as described in Sect. 5.4.2 until an
$r$-packet is emitted. On the other hand, if the result is a $k$-packet, it
converts in situ into an $r$-packet as described in Sect. 5.4.3.

\subsection{New direction cosine}   

Re-emission following a physical event is assumed to occur
isotropically in the matter frame, so the new direction cosine in every
case is $\mu = 1-2z$.  
For electron scattering, this is an approximation. But isotropic
emission is exact for
thermal processes - i.e., for f-f and f-b emissions. It is also exact for
the Sobolev treatment of b-b emission
in a homologously expanding flow since there is then no
kinematically-preferred direction. 

	If this technique is applied to a spherically-symmetric stellar wind,
the expansion is no longer homologous and so b-b emission is angle dependent.
Specifically, the pdf to be sampled for the new direction cosine is
$ \propto I_{\nu_{k}}^{r}(\mu)/[\mu^{2} dv/dr + (1 - \mu^{2}) v/r]$, where
$ I_{\nu_{k}}^{r}(\mu)$ is the intensity in the line's far-red wing
of the radiation emitted by the line (Castor 1970; Lucy 1971). 

\subsection{New rest energy} 
  
If $\epsilon_{R}$ is the packet's
rest energy during its free flight prior to event $n+1$, then, at the event, 
$\epsilon_{\nu} = \epsilon_{R} (1 - \mu_{1} v/c)$, where $v$ is the velocity
at that location and
$\mu_{1}$ is the incident direction cosine. Now, at every event, the incident
and emergent $r$-packets have the same energy $\epsilon_{\nu}$. Accordingly,
the packet's rest energy along its free flight following event $n+1$
is $\epsilon_{R} =\epsilon_{\nu}/(1 - \mu v/c) $, where  $\mu$ is the
new direction cosine.

	With the assignment of $\nu$ and $\mu$ and the updating of
$\epsilon_{R}$, the treatment of event 
$n+1$ is complete. The calculation therefore now returns to Sect. 5.1
to initiate the search for event $n+2$.
  
\section{Improved solution}   

The trajectories of ${\cal N}$ individual $r$-packets computed as
described in Sect. 5 collectively constitute the MC estimate of the
radiation field in $D$.
The next challenge is to use it to derive improved values of $n_{i}$ and
$T$ for each of the ${\cal M}$ constant-density shells.

\subsection{Improved scale factor and boundary temperature}   

The MC estimate for the luminosity of the SN is
\begin{equation}
 L(\infty) = \frac{\epsilon_{0}}{\Delta t} \sum 
                                  \frac{\epsilon_{R}}{\epsilon_{0}}
\end{equation}
where the summation is over the $r$-packets that escape to $\infty$.

	From this equation, we derive an improved scale factor
$\epsilon_{0}/\Delta t$ by requiring that $L(\infty)$ has its specified value 
- see Sect. 2.2. The improved value of $T_{b}$ then follows from Eq. (21). 

	This improved scale factor is used below in computing the radiative
rates needed for the statistical and thermal equilibrium calculations. The
improved $T_{b}$ is used to determine
the initial frequency distribution of $r$-packets (Sect. 4.1)  
at the start of the next iteration.

\subsection{Monte Carlo estimators}   

	In order to use the MC radiation field to improve the values of
$n_{i}$ and $T$, estimators are needed for the 
radiative rate coefficients. Fortunately, efficient estimators can be
constructed with a procedure given in an earlier paper (Lucy 1999a).

	In the context now of a moving atmosphere, the energy-density argument
of Lucy (1999a) applied to a local co-moving frame yields the formula  
\begin{equation}
 J_{\nu}\: d \nu = \frac{1}{4 \pi} \frac{\epsilon_{0}}{\Delta t} \:
               \frac{1}{V}  \:
       \sum_{d \nu}  \frac{\epsilon_{\nu}}{\epsilon_{0}} ds
\end{equation}
where the summation is over all trajectory segments $ds$  
in $V$ such that the $r$-packets' frequencies $\nu (s)$ 
fall in the interval $(\nu, \nu + d \nu)$. This equation is the building
block that allows us to construct the required estimators.

	For this problem, $V$ in Eq. (43) is the volume of one of the
spherical shells into which the envelope is discretized (Sect. 3.2). But
the estimators derived in this section are quite general: no particular
geometry or symmetry is assumed.  

	If we use Eq.(43) to derive an estimator
for the photoionization coefficient given by Eq. (16), the result is
a summation, with the summed quantity being 
the integral of $\epsilon_{\nu} a_{i \kappa}(\nu) / h \nu$
over the pathlength $\Delta s$ between consecutive events. But since every
pathlength is confined to a single shell, $\Delta s \rightarrow 0$ as
${\cal M} \rightarrow \infty$ and so $\Delta s$ may be treated as a small
quantity.
This then implies that each $\Delta s$-integral can be approximated by a
one-point quadrature. With this approximation, the estimator for the
photoionization coefficient for level $i$ is 
\begin{equation}
 \gamma_{i} = \frac{\epsilon_{0}}{\Delta t} \:
               \frac{1}{V}  \:
       \sum_{\nu > \nu_{i}}  \frac{\epsilon_{\nu}}{\epsilon_{0}} \: \Delta s 
                 \:  \frac{a_{i \kappa}(\nu)}{h \nu}
\end{equation}
where $\nu$ is the packet's frequency at the mid-point of $\Delta s$ and
the summation is over all $r$-packets in $V$ with $\nu > \nu_{i}$.

	The efficiency of this estimator derives from not being
restricted to $r$-packets that actually cause photoionizations. 
Thus its precision greatly exceeds that of an estimator based simply on
counting photoionizations in $V$.
In particular, this estimator returns a non-zero estimate
of $\gamma_{i}$ even if no $r$-packets cause photoionizations from level $i$
in a given
shell, as commonly happens for the higher, less populated levels and in
the outer, less dense shells. 

	Applying the same procedure now to Eq. (15), we obtain
\begin{equation}
 \alpha^{st}_{i} = \Phi_{i \kappa} \: \frac{\epsilon_{0}}{\Delta t} \:
               \frac{1}{V} \:
     \sum_{\nu > \nu_{i}}  \frac{\epsilon_{\nu}}{\epsilon_{0}} \: \Delta s \:
                   \frac{a_{i \kappa}(\nu)}{h \nu} \: 
                      e^{-h \nu / kT}
\end{equation}
as the corresponding estimator for the rate coefficient of
stimulated recombinations to level $i$, with $\nu$ again evaluated at the
mid-point of $\Delta s$. This estimator's efficiency
is a consequence of not being limited to $r$-packets that actually cause
stimulated recombinations.
	
	The above estimators refer to transitions to and from the continuum.
But statistical equilibrium also depends on the radiative rates
of b-b transitions - see Eq.(10). Thus we require an estimator for
$J_{ji}^{b}$, the mean intensity at frequency $\nu_{ji}^{+}$
 - i.e. in the far blue wing of the transition $j \rightarrow i$. This can
also be derived from Eq.(43).

	For homologous expansion, there is no kinematically-preferred
direction, and so
the rate at which the frequency $\nu(s)$ of an $r$-packet
in free flight decreases with distance $s$ is independent of its direction of
propagation.
To the accuracy of this calculation - see Sect. 3.3, this rate is 
$d \nu/ds =-\nu_{R}/ct$. This equation gives the length $ds$ of the
trajectory segment within which $\nu$ is in the interval
$(\nu_{ji}, \nu_{ji} + d \nu)$. Substituting this value of $ds$ into Eq. (43),
we obtain
\begin{equation}
 J_{ji}^{b} =    \frac{\lambda_{ji} t}{4 \pi} \:
       \frac{\epsilon_{0}}{\Delta t} \: \frac{1}{V}  \:
 \sum \frac{\epsilon_{ji}}{\epsilon_{0}} \: \frac{\nu_{ji}}{\nu_{R}}
\end{equation}
as an estimator for the mean intensity at $\nu_{ji}^{+}$. In this formula,
$\epsilon_{ji}$ is the packet's energy at
the point where $\nu = \nu_{ji}$, and the
summation is over all $r$-packets in $V$ that come into resonance with the
transition $j \rightarrow i$.

	The efficiency of this estimator
derives from it not being restricted to $r$-packets that actually undergo
the b-b transition $j \rightarrow i$. Accordingly, a non-zero estimate is
usually returned even when no such transitions occur in $V$. Evidently, the
estimate is zero only if no $r$-packets in $V$ happened to come into resonance
during the MC calculation. To avoid such spurious zero estimates for
$J_{ji}^{b}$, a minimum value of ${\cal N}$ is required (Lucy 1999b).    

	The above estimators refer only to the radiative rates that enter the
equations of statistical equilibrium. Further quantities are required by
the equation of thermal equilibrium. These include the coefficients
$\gamma_{i}^{E}$ and $\alpha^{E, st}_{i}$ defined in Sect. 3.6. As is   
evident from their definitions, estimators for these
quantities are obtained from those derived above for $\gamma_{i}$ and
$\alpha^{st}_{i}$ simply by changing $h \nu$ to $h \nu_{i}$ in the factors
$a_{i \kappa}(\nu)/h \nu$ - see Eqs. (44) and (45).  

	Finally, an estimate is also needed for the heating rate 
due to f-f absorptions,
\begin{equation}
 {\cal H}^{ff} = 4 \pi \int_{0}^{\infty} k^{ff}_{\nu} \rho \: 
                                                 J_{\nu} \:  d \nu
\end{equation}
If we separate out the dependence on level population by writing
${\cal H}^{ff} = h_{\kappa}^{ff} n_{\kappa}$ and then
again apply Eq. (43), the estimator for the coefficient is 
\begin{equation}
  h_{\kappa}^{ff} =n_{e} \frac{\epsilon_{0}}{\Delta t} \:
               \frac{1}{V}  \:
       \sum  \frac{\epsilon_{\nu}}{\epsilon_{0}} \: \Delta s 
                 \: \frac {k^{ff}_{\nu} \rho}{n_{\kappa}n_{e}}  
\end{equation}
where the summation is over all pathlengths in $V$ and
$\nu$ is evaluated at their mid-points.

	The summations in the above estimators are accumulated in the course 
of the MC simulation and are the only quantities concerning
the internal radiation field that need to be stored. When the trajectories
of all ${\cal N}$ packets have been computed, the scale factor 
$\epsilon_{0}/\Delta t$ is determined as described in Sect. 6.1 and 
then used to convert the summations into the required rate coefficients.

	Although the accuracy of these various estimators remains to be 
demonstrated, it is clear from their derivation that they are consistent.
In other words, as ${\cal N}  \rightarrow \infty$ {\em and}
${\cal M} \rightarrow \infty$,
they converge to the quantities being estimated.

\subsection{Statistical equilibrium}   

	The required solution is in statistical and thermal
equilibrium. Accordingly, following each MC calculation,
the equations representing both of these equilibria are used
simultaneously to derive improved values of $n_{i}$ and $T$ throughout the
envelope. Nevertheless, it is convenient to discuss these equilibria
separately.  

	The equation of statistical equilibrium for level $i$ can be
written in the form
\begin{equation}
 \Lambda_{\ell i} n_{\ell} -  (\Lambda_{i \ell} + \Lambda_{i u}) n_{i}
                   + \Lambda_{u i}n_{u}=0
\end{equation}
where $\Lambda_{ij}$ denotes the total rate coefficient for the transition
$i \rightarrow j$. Specifically, for transitions between bound levels, we
have [cf. Eqs (9) and (10)]
\begin{equation}
 \Lambda_{ji} = B_{ji} \beta_{ji} J_{ji}^{b} + q_{ji} n_{e}
\end{equation}
for excitations $j \rightarrow i$, and
\begin{equation}
 \Lambda_{ij} =  A_{ij}\beta_{ji} + B_{ij} \beta_{ji} J_{ji}^{b} + q_{ij}n_{e}
\end{equation}
for de-excitations $i \rightarrow j$. In addition, for transitions coupling
bound levels to the continuum, we have
\begin{equation}
 \Lambda_{i \kappa} = \gamma_{i} + q_{i \kappa} n_{e} 
\end{equation}
for ionizations  $i \rightarrow \kappa$, and
\begin{equation}
 \Lambda_{\kappa i} = (\alpha_{i}^{sp}+\alpha_{i}^{st})n_{e} +
 q_{\kappa i} n_{e}^{2}  
\end{equation}
for recombinations $\kappa \rightarrow i$.

	Eq. (49) applies to all levels $i = 1 \ldots \kappa$. But since every 
source term is a loss term for another level, the sum of the left hand sides
is zero, which implies that one of the equations is redundant. This leaves
$\kappa - 1$ equations in the $\kappa$ unknown level populations $n_{i}$. To
make the system determinate, we impose the constraint
\begin{equation}
  n_{1}+n_{2}+ \cdots + n_{\kappa} = n_{H} 
\end{equation}
where $n_{H}$ is the known number density of H atoms - see Sect. 2.1.

	The coefficients $\Lambda_{ij}$ depend on $n_{\kappa} (= n_{e})$ 
and on the $n_{i}$ of bound levels through
the escape probabilities $\beta_{ji}$.
Accordingly, the system of equations defined by
Eqs. (49) and (54) is non-linear. But fortunately the system can be solved
iteratively with a 
standard package for linear equations. Thus the coefficients are evaluated
with the current estimates of $n_{i}$ and then the system solved for
improved estimates. These are then used to recompute the $\beta_{ji}$ and
hence the $\Lambda_{ij}$ and the process repeated until convergence is
achieved.

	However, at this point a limitation in the MC approach to NLTE
problems becomes apparent. The problem arises for the populations $n_{i}$
of closely-spaced levels near the continuum. Monte Carlo sampling errors  
in the estimated rates of the processes populating such levels can give rise
to level inversions, and these then prevent convergence of the above back
substitutions. To overcome this, the following stabilizing device is adopted:
for each level $j$,
the $n_{i}$ of a higher level $i$ is replaced by $n_{c} = n_{j}g_{i}/g_{j}$
if $n_{i} > n_{c}$.

	Because this problem only arises for high levels in the outermost
shells, its impact on
the MC code's prediction of the SN's line- and continuum spectrum from the
UV to the mid-IR is totally negligible. Moreover, as expected, the occurrence
of these spurious inversions decreases
as the size of the MC simulation increases. For example,
with ${\cal N} = 2 \times 10^{4}$ in an envelope stratified into
${\cal M} = 20$
shells, inversions occurred in 13 shells starting at $m = 2$ and affecting
levels as low as $i=6$. With ${\cal N}$ increased to $5 \times 10^{4}$,
inversions occur in only 5 shells starting at $m = 8$ and the lowest
affected level is $i = 8$. With a modest further increase to 
${\cal N} \ga 10^{5}$, inversions seldom occur. Nevertheless, problems
requiring accurate population ratios for closely-spaced levels are perhaps
best not tackled with MC methods. 

\subsection{Thermal equilibrium}   

In addition to statistical equilibrium, the required NLTE solution
must also be in thermal equilibrium. This latter equilibrium is simply the
condition that throughout the envelope
\begin{equation}
 {\cal G} =  {\cal H} - {\cal C} = 0
\end{equation}
where ${\cal H}$ and ${\cal C}$ are the total heating and cooling rates,
respectively. Given the physical processes included in this
calculation, we have
\begin{equation}
  {\cal H} = {\cal H}^{bf} +  {\cal H}^{ff} + {\cal H}^{c \ell}
\end{equation}
and 
\begin{equation}
  {\cal C} = {\cal C}^{fb} +  {\cal C}^{ff} + {\cal C}^{c \ell}
\end{equation}

	In terms of the modified photoionization coefficient defined in
Sect. 3.6, the rate at which b-f transitions convert radiant energy into
thermal energy is 
\begin{equation}
  {\cal H}^{bf} = \sum_{1}^{\kappa -1} \:  h_{i}^{bf} n_{i} \;\;\;\; 
\mbox{with}
 \;\;\;\;         h_{i}^{bf} =  (\gamma_{i}^{E} - \gamma_{i}) \: h \nu_{i}
\end{equation}
The corresponding rate at which f-b transitions convert thermal energy 
into radiant energy is
\begin{equation}
  {\cal C}^{fb} = c_{\kappa}^{fb} n_{\kappa}   \;\;\;\; \mbox{with} \;\;\;\;
c_{\kappa}^{fb} = n_{e}\sum_{1}^{\kappa -1} \: (\alpha_{i}^{E} - \alpha_{i})
 \: h \nu_{i} 
\end{equation}

	The heating and cooling rates ${\cal H}^{ff}$ and ${\cal C}^{ff}$
due to f-f processes have previously been
defined in Eqs. (47) and (32). The cooling rate ${\cal C}^{c \ell}$ due to
collisional excitations is given by Eq. (33), and the corresponding
heating rate ${\cal H}^{c \ell}$ due to de-excitations is
computed similarly. 

	When the MC model of the internal radiation field has been computed,
the current values of a shell's level populations and temperature,
$\bar{n}_{i}$ and $\bar{T}$, do not in general correspond to thermal
equilibrium - i.e.,
$ \bar{{\cal G}} \neq 0$ . To eliminate this residual,
we apply corrections $\delta n_{i}$ and $\delta T$ that satisfy the linearized
version of Eq.(55), 
\begin{equation}
  \bar{{\cal G}} + \frac{\partial{\cal G}}{\partial n_{i}} \delta n_{i}
             +\frac{\partial{\cal G}}{\partial T} \delta T = 0
\end{equation}
But a more convenient form of this equation follows from noting that
each level's contribution to the net heating rate can be expressed in terms
of heating
($h_{i}$) and cooling ($c_{i}$) coefficients  - see Eqs (48),(58) and (59) -
so that $G = \sum (h_{i}-c_{i})n_{i}$.  
With this substitution, Eq.(60) 
becomes  
\begin{equation}
 \sum_{1}^{\kappa} (h_{i}-c_{i})n_{i}
             +\frac{\partial{\cal G}}{\partial T} \delta T = 0
\end{equation}
where $\partial {\cal G}/ \partial T$ is computed by numerical 
differencing, and
$n_{i} = \bar {n}_{i} + \delta n_{i}$ is the improved level population.

	Eq. (61) is one equation in $\kappa +1$ unknowns. But if
we add it to Eqs.(49) and (54), we have
$\kappa +1$ equations in the $\kappa +1$ unknowns $n_{i}$ and $\delta T$.
This combined system of equations, representing statistical {\em and} thermal
equilibrium, is solved
by repeated back substitution as described in
Sect. 6.3. With each back substitution, the coefficients $\Lambda_{ij}$,
$h_{i}$, $c_{i}$ and  $\partial{\cal G}/ \partial T$ are recomputed because
of their dependences on  
$n_{i}$, $n_{e}$ and $T$.

	This simple iteration technique for finding 
$n_{i}$ and $\delta T$ must often start far from the final solution and can
then fail due to large initial corrections.
To avoid such failures, an undercorrection factor
$u = 0.8$ is applied to the corrections $\delta n_{i}$ and $\delta T$,
and the temperature change is further limited to $\pm \Delta T$, with
$\Delta T = 200$K. Moreover, if
the magnitude of the correction vector $\delta n_{i}$ increases from one
iteration to the next, the factor $u$ and the bound $\Delta T$ are both
decreased. With these precautions, back-substitution proves
to be a robust technique for solving the equations of statistical and thermal
equilibrium.

	The iteration technique also fails when sampling errors'
contribution to ${\cal H} - {\cal C}$ itself implies a large correction
$\delta T$. The only remedy in this case is to increase ${\cal N}$.

	The iterations are deemed to have converged when
$|\delta T| < 1$K  {\em and}
$\overline{|\delta n_{i}|/n_{i}} < 10^{-5}$, where the averaging  
is over levels $i=1-5$
because of the occasional spurious inversions of high levels (Sect. 6.3).

	When the iterations have converged to this accuracy, the typical
residuals for each shell are $|{\cal H}-{\cal C}| / {\cal C}| < 10^{-4}$ and
similarly for the agreement between the source and sink terms, 
$\Lambda_{\ell i} n_{\ell} + \Lambda_{u i}n_{u}$ and
$(\Lambda_{i \ell} + \Lambda_{i u}) n_{i}$, for all levels $i$ in Eq.(49).
However,
when the $n_{i}$ of a high level has been stabilized to avoid a spurious
inversion, the fractional departure from statistical equilibrium rises to
$\sim 10^{-3}$.

\subsection{Outer iteration}   

During the above (inner) iterations to find $n_{i}$ and $T$,
the pathlength summations of Sect. 6.2 are of necessity held fixed.
In effect, therefore, the matter in each shell is brought 
into equilibrium with a {\em fixed} radiation field. 
Accordingly, if the MC radiation field were now to be recomputed with the
improved 
$n_{i}$ and $T$, these summations would change, and so 
statistical and thermal equilibrium would again be violated.
Evidently, an outer iterative loop is necessary if 
convergence to the NLTE solution is to be achieved.

	The scheme adopted for these outer iterations is simply to
repeatedly input the updated $n_{i}$ and $T$ from Sect. 6.4 into the MC
calculation of Sect. 5. These outer iterations are continued until the 
changes in $n_{i}$ {\em and} $T$ from one outer iteration to the next 
can be attributed to MC sampling errors.

	The novelty of this iterative technique is that, despite
well-documented convergence failures in various earlier applications
(see, e.g., Mihalas 1978), we are relying
on the $\Lambda$-iteration device of repeatedly
bringing matter into statistical and thermal equilibrium with the just-
updated MC radiation field. Indeed, these $\Lambda$-iterations would also
fail to
converge {\em if} the radiation field were obtained by solving the
Equation of Radiative Transfer (RTE).
But in this MC scheme, the radiation field is subject at every iteration
to constraints that other iterative schemes only aim to achieve
asymptotically.
Thus, even while $n_{i}$ and $T$ still depart from their equilibrium values,
the constraint of radiative equilibrium in the matter frame is 
obeyed rigorously. 
Moreover, the MC transition probabilities approximately
incorporate the constraints that statistical equilibrium imposes on the
frequency redistribution of absorbed radiant energy. 
In effect, therefore, the scheme introduces a {\em constrained}
$\Lambda$-operator that, for sufficiently large ${\cal N}$,
generates at every iteration a radiation field
that is closer to the converged solution than would be the radiation field
generated by the RTE. The convergence of these constrained
$\Lambda$-iterations is illustrated in Sect. 7.2.

\section{Numerical results}   

	In Sects. 3-6, a detailed description has been given of how the MC
model of the internal radiation field is derived and how it
is used to improve level populations and temperatures throughout
the envelope. But whether such improvements will actually result in 
convergence to the NLTE solution remains to be demonstrated.

\subsection{Accuracy of estimators}   

Crucial to the success or otherwise of this technique is the
accuracy of the estimators developed in Sect. 6.2.
Even with consistent estimators, the technique would have little current
interest if adequate accuracy requires simulations with 
impossibly large ${\cal N}$. On the other hand, if accurate rates are
achievable with feasible values of ${\cal N}$, then the coefficients in,
and therefore the solutions of, the equations of statistical and thermal
equilibrium will differ little from those derivable with a standard transfer
calculation.

	An obvious accuracy test would be to obtain an accurate solution of a
NLTE problem using a conventional code
and then examine the convergence of MC solutions of the same problem 
as ${\cal M}$ and ${\cal N}$ increase.
But a simpler procedure is followed here: the special case of free-streaming 
$r$-packets allows
values of the various coefficients obtained with the MC estimators to
be compared with values obtained by numerical integration.

	The MC code is easily modified so that $r$-packets 
propagate without interaction.
Each packet is then emitted as before from the moving lower
boundary but thereafter conserves its $\epsilon_{R}$ and $\nu_{R}$, while its
$\epsilon_{\nu}$ and $\nu$ decrease along the trajectory 
because of the differential expansion. 

	A further modification made for this free-streaming test is to
omit stratified sampling of the $r$-packets' frequencies at the lower
boundary - see Sect. 4.1.
Thus, the ${\cal N}$ bins in $\ell n \: \nu$ are here randomly not
uniformly sampled. By thus not benefiting from stratified sampling, the
results should be more typical of circumstances where the radiation field
is created within $D$ rather than emitted by an artificial boundary condition.
 
	At radius $r$, the conical radiation field modelled
by this free-streaming MC calculation has specific intensity
$I_{\nu}(\mu) =  B_{\nu'}(T_{b})$, where
$\nu'(\mu)$ is the frequency at $r = R$ of a photon that
reaches $r$ with frequency $\nu$ and direction cosine $\mu$. From this
expression for $I_{\nu}$, the mean intensity $J_{\nu}(r)$ can be computed to
arbitrary accuracy by numerical integration. Such calculations carried out
at fequencies $\nu_{ji}$ and at the mean radii $\bar{r}_{m}$ then allow the
estimator $J_{ji}^{b}$ given by
Eq. (46) to be tested. Similarly, $J_{\nu}(\bar{r}_{m})$ can be computed
for a grid of frequencies and then used to obtain
accurate values for the coefficients $\gamma_{i}$, $\gamma_{i}^{E}$,
$\alpha_{i}^{st}$, $\alpha_{i}^{E, st}$ and $h_{\kappa}^{ff}$ by numerical
integration, thereby testing their MC estimators from Sect. 6.2.

	These comparisons, carried out for ${\cal M} = 10 - 100$ and 
${\cal N} = 10^{4} - 10^{6}$, support the assertion of Sect. 6.2 that these
estimators are consistent. Thus numerical experiments indicate that the
errors of the MC estimates $\rightarrow 0$ as    
${\cal M}$ {\em and} ${\cal N \rightarrow \infty}$, with no suggestion of
bias. But if ${\cal M}$ is fixed while ${\cal N \rightarrow \infty}$,
the estimators' sampling errors $\rightarrow 0$ but biases due to 
discretization errors remain. On the other hand, if ${\cal N}$ is fixed while
${\cal M \rightarrow \infty}$, biases $\rightarrow 0$ but sampling errors
become large, especially for the mean intensities $J_{ji}^{b}$. 

	In addition to confirming the estimators' consistency, this
special case also suggests that acceptable accuracies are indeed achieved
with feasible values of $\cal N$. For example, 
with ${\cal M} = 20$ and ${\cal N} = 10^{5}$, the MC free-streaming 
simulation requires only about 3 min computer time and yet gives reasonably 
accurate coefficients. Thus, in one such simulation, the means of the
fractional and of the absolute fractional errors of the coefficients
$\gamma_{i}$ for $i = 2, \ldots ,\kappa-1$ and $m = 1, \ldots ,{\cal M}$ are 
$2.8 \times 10^{-3}$ and $6.1 \times 10^{-3}$, respectively. These
gratifyingly small errors are a consequence of the large number of
$r$-packets contributing to the summation in Eq. (44) and confirm the
efficiency of estimators constructed from Eq. (43).

	The quantities least accurately estimated are the $J_{ji}^{b}$.
Thus, in the above simulation, the mean fractional and absolute fractional
errors of the $J_{ji}^{b}$ for the Balmer lines in all shells are
$0.4 \times 10^{-2}$ and $12.6 \times 10^{-2}$, respectively. In this case,
sampling errors clearly dominate over any systematic bias due to  
discretization errors.
This is no suprise since the summation in Eq. (46) includes only those
$r$-packets that are red-shifted into resonance with the transition 
$j \rightarrow i$. Accordingly, even with ${\cal N} = 10^{5}$, the number
of packets contributing to the summation for a particular $J_{ji}^{b}$ 
is not large. For example, in the case of the Balmer lines, the 
number is $\sim 35$ at $m = 1$ rising to $\sim 300$ at $m = {\cal M}$. 

	These relatively large sampling errors for the $J_{ji}^{b}$
are the main source of error for the coefficients in the
equations of statistical equilibrium.
Of course, for a typical level, there are several source terms, some 
not even subject to sampling errors; and so the fractional sampling
error of the total input rate will be correpondingly reduced. Nevertheless,
for a level populated predominately by a single radiative b-b transtion,
its $n_{i}$
will directly reflect the sampling error of the relevant
$J_{ji}^{b}$. But since sampling errors differ from one iteration to
the next, levels suffering this sensitivity will display fluctuating values
of $n_{i}$ and so should not escape notice.

\subsection{Convergence of temperature profile}   

	To investigate convergence to the NLTE solution, the values of
$T$ and $n_{i}$ throughout the envelope were monitored for 20 outer 
iterations in a calculation with ${\cal M} = 20$ shells and   
${\cal N} = 4 \times 10^{5}$ packets. The envelope's parameters are as
specified in Sect. 2.2, and the calculation is initiated as described in
Sect. 3.4. Thus, for the first iteration, the MC transition probabilities
are computed  with radiative and collisional rates corresponding to the
initial analytical
model of the radiation field -Eq. (4)- and to the inital guesses for
$n_{i}$ and $T$. For all subsequent iterations,
the MC transition probabilities are computed from the MC radiation field
via the MC estimators of Sect. 6.2, and the $n_{i}$ and $T$ are the 
updated values resulting from bringing the matter into statistical and
thermal equilibrium with the MC radiation field. 

	Figure 1 shows how the envelope's temperature stratification
evolves from the initial $T = 5000$K as the iterations proceed.
The corrections are seen to
be substantial for iterations 1-4 and negligibly
small thereafter. Thus the scheme appears to converge at the fifth iteration,
with the changes introduced at iterations 6-20 presumably being
random fluctuations about the exact NLTE solution. Such fluctuations are
expected because of different sampling errors in successive realisations of
the MC radiation field.

\begin{figure}
\vspace{8.2cm}
\includegraphics{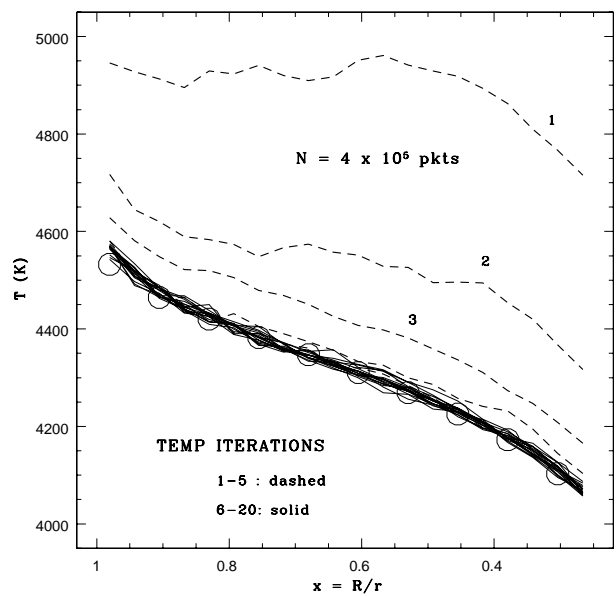}
\caption{Convergence of constrained $\Lambda$-iterations. Results of 20
 temperature iterations starting from an isothermal stratification with
 $T = 5000$K are plotted for a simulation with ${\cal N} = 4 \times 10^{5}$
 packets. The solution for the conical radiation field is plotted as open 
 circles.}
\end{figure}

	To support this claim that the scheme has indeed achieved rapid
convergence
to the close neighbourhood of the exact NLTE solution, a separate
NLTE calculation has been carried out with line formation again treated
in the Sobolev approximation but now with the continuum radiation field  
approximated by the free-streaming model described in Sect. 7.1. With the
radiation field thus prescribed, the determination of $n_{i}$ and $T$ at
radius $r$ from the equations of statistical and thermal equilibrium is
a one-point problem. The coefficients in these equations are calculated 
with accurate numerical integrations as for the accuracy tests of Sect. 7.1
and then $n_{i}$ and $T$ obtained with the back-substitution iterations of 
Sect. 6.4. This one-point calculation has been carried out at several radii,
with the results plotted as open circles in Fig. 1. The agreement with
the MC solutions for iterations 6-20 is excellent.  

	Given that the scheme has indeed converged and that the remaining
temperature
fluctations are therefore due to sampling errors, the next question is
whether such fluctuations are unacceptably large. Assuming convergence at
iteration 6, we derive an accurate estimate $\overline{T}$ of the exact $T$
for each shell by averaging over iterations 6-20. The standard deviation
of the fluctuations $T-\overline{T}$ can then be computed for each shell as
well as a consolidated value $\sigma_{T}$ for the entire envelope. For this
simulation with ${\cal N} = 4 \times 10^{5}$, we find that 
$\sigma_{T} = 9.0$K. Temperature uncertainties of this magnitude are
admittedly large
compared to
the nominal precision - often $< 0.01K$ - achieved by conventional NLTE codes.
But in view of the parametric, geometrical
and kinematic uncertainties associated with specifying the structure of
any celestial object, errors of $\sim 10$K are completely inconsequential.

	The amplitude of the fluctuations following convergence depend of
course on ${\cal N}$. Repeating the above simulation with 
${\cal N} = 10^{5}$ and again averaging over iterations 6-20,
we find that $\sigma_{T}$ increases to $16.2$K, consistent with the expected
scaling $\sigma_{T}  \propto {\cal N}^{-1/2}$.

\subsection{Convergence of level populations}   

	The convergence behaviour of the $n_{i}$ for the same 20 iterations
of the above model with ${\cal M} = 20$ shells and   
${\cal N} = 4 \times 10^{5}$ packets is shown in Fig. 2 for shell $m = 10$.
Again we see substantial corrections in the first few iterations followed
by fluctations of moderate amplitude about the scheme's solution in the
limit $ {\cal N} \rightarrow \infty$.

	As for the temperature iterations, the MC solution is compared to
the accurate NLTE solution for the conical radiation field of Sect. 7.1.
The agreement is excellent for low levels and for the continuum ($\kappa$)
but small discrepancies are evident for high levels. This probably reflects
a small difference in the two calculations due to the device used in the
MC calculation to avoid negative probabilities when $b_{\kappa} > b_{i}$
 - see Sect. 4.3.1.

\begin{figure}
\vspace{8.2cm}
\includegraphics{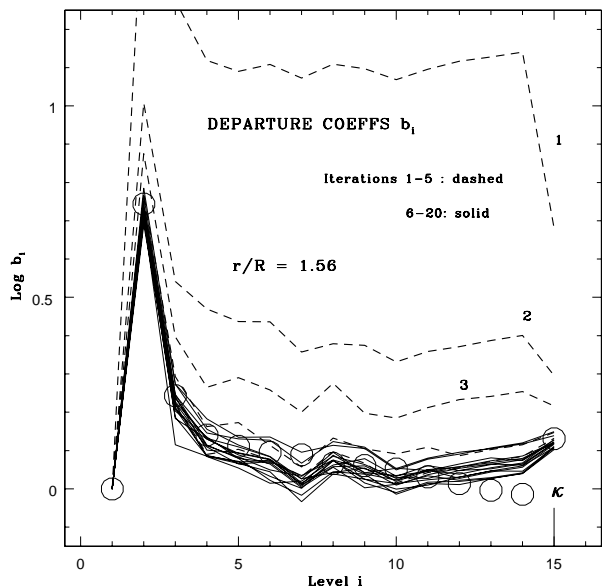}
\caption{Convergence of constrained $\Lambda$-iterations. Results of 20
level population iterations starting from Saha-Boltzmann populations at
$T = 5000$K are plotted for the mass shell $m = 10$ in a simulation with
${\cal N} = 4 \times 10^{5}$
packets. The solution for the conical radiation field is plotted as open 
circles.}
\end{figure}

	A further test of the accuracy of the converged $n_{i}$ is presented
in Fig. 3. In this diagram, the variation with radius of the departure
coefficients for several levels are compared to the corresponding values 
for the conical radiation field. The agreement is seen to be satisfactory.

\begin{figure}
\vspace{8.2cm}
\includegraphics{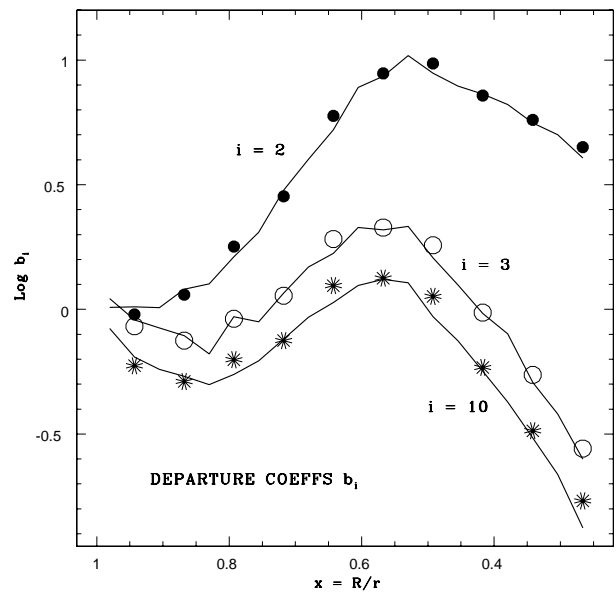}
\caption{Test of departure coefficients $b_{i}$. Results obtained at
iteration 20 in Monte Carlo calculation with ${\cal N} = 4 \times 10^{5}$
packets are plotted as solid lines for the indicated levels.
The solutions for the conical radiation field are plotted as filled circles
($i=2$), open circles ($i=3$) and asterisks ($i=10$).}
\end{figure}

	The behaviour of the H level populations in this SN envelope are
qualitatively similar to those of the hydrogenic ion He$^{+}$ in W-R winds
(Hillier 1987). In particular, photoionizations predominantly occur from
level 2, which 
behaves like a ground state, becoming overpopulated in the outer layers due
to the dilution of the ionizing continuum shortward of $\lambda$3646 \AA\
and to the trapping of Lyman $\alpha$ photons.

	Note that, because most H atoms are in the ground state for both the
NLTE and the LTE solutions, the departure coefficient $b_{1}$ is almost
exactly $= 1$. Accordingly, the values $b_{2}$ plotted in Fig. 3 directly
indicate the decoupling of level $n = 2$ from the ground state.

\subsection{Consistency}   

Given that the macro-atom formalism has now been extended to include 
interactions with the thermal pool, it is of interest to test this extension 
by repeating the consistency test of level emissivities described for H
in Sect. 5.2 of Paper I.

	When the outer iterations have converged, the level populations,
temperature and radiative rates appropriate to statistical {\em and } 
thermal equilibrium are known for each mass shell. From this 
information, the rates at which radiant energy is being converted to other
forms can be computed for each of the processes operating. Specifically,
these are: conversions into excitation energy by b-b absorptions to levels
$2, \ldots , \kappa-1$; conversion into ionization energy by b-f absorptions;
and conversions into thermal energy by b-f and f-f absorptions. These last
two processes, labelled $\kappa+1$ and $\kappa +2$, were not included in
Paper I, which treated only statistical equilibrium.    

	From these conversion rates, a Monte Carlo test of the MC transition
probabilities is carried out for each mass shell as follows: $ N = 10^{6}$
equal energy $r$-packets are assigned to these absorption
channels in
proportion to the computed conversion rates. Those assigned to processes
$2, \ldots ,i, \ldots  ,\kappa$ activate a macro-atom to state $i$, while
those assigned to processes $\kappa+1$ and $\kappa +2$ result in the creation
of a $k$-packet. Application of the MC transition probabilities then
results eventually in the emission of an $r$-packet as described in Sect. 4.
Moreover, each emitted $r$-packet belongs to one of the $\kappa + 1$
emission processes representing the inverse of the above absorption
processes. Accordingly, when all $N$ packets have been absorbed and
re-emitted, we have experimental values $N_{i}$ for each of the $\kappa + 1$
emission processes.

	But the rates at which excitation, ionization and thermal energy
is being converted in a given shell into radiant energy can be directly
computed from the 
known level populations and temperature - i.e., from the rates of the
various b-b, f-b and f-f processes. Thus, we can predict that
$N^{*}_{i}$ of the $N$ packets should have been emitted by process $i$.

	In Paper I, Fig. 4, the agreement of $N_{i}$ and $N^{*}_{i}$ was
shown graphically. Here, more rigorously, we compute the values of 
$\chi^{2} = \sum_{i} (N_{i}-N^{*}_{i})^{2}/N^{*}_{i}$ for each mass shell.
Since there are 16 emission channels, we expect the experimental values of 
$\chi^{2}$ to be distributed as $\chi^{2}_{\nu}$ with $\nu = 15$ degrees of
freedom. Thus, 50 percent of the values should be $< 14.34$ and 
95 percent $< 25.00$ 

The results obtained after the 20th iteration for the model with
${\cal M} = 20$ discussed in
Sect. 7.2 are as follows: the 20 values of $\chi^{2}$ range from 7.87 to
34.22, with $6 < 14.34$ and $3 > 25.00$.

	The outcome of this test is that the experimental values of
$\chi^{2}$ are slightly biased to higher values than predicted by the 
$\chi^{2}_{\nu}$ distribution. However, given that the underlying model
has departures from strict statistical and thermal equilibrium because
of its own MC sampling errors, these results are in fact a strong 
confirmation that
the extension of the macro-atom formalism to include interactions with the
thermal pool has been carried out correctly.    

	As a further test of this conclusion, values of $\chi^{2}$ were
computed with the summation restricted to the three channels containing
$r$-packets emitted by f-b and f-f processes. In this case, the values for
all shells are $ < 7.81$, the 95 percent confidence limit for $\nu = 3$
degrees of freedom. This eliminates the possibility that the above upward
bias is due to poor predictions for the continuum channels. 
 
\subsection{Monte Carlo spectrum}   

	When the outer iteration loop has converged, the $r$-packets that
escaped to $\infty$ during the last iteration provide a crude estimate of the
SN's emergent spectrum. This is obtained by simply allocating each escaping
packet's rest energy $\epsilon_{R}$ to its appropriate bin in a grid of rest
frequencies. This crude spectrum is shown in Fig. 4 for the above model with
${\cal N} = 4 \times 10^{5}$ packets. In this figure, luminosity density is
plotted against vacuum wavelength in the optical domain and, despite sampling
errors, the P Cygni line profiles of H$\alpha$ and H$\beta$ are clearly
seen. 

	If this MC code were intended for use in interpreting observational
data, the crude MC spectrum shown in Fig. 4 would not be compared with the
observed spectrum. Instead, a far less noisy theoretical spectrum would be
computed by following a procedure described earlier (Lucy 1999b).
Specifically, estimates of line- and continuum source functions can be
derived from the same 
MC simulation and used to compute a high quality spectrum from the
formal integral for the emergent intensity as a function of impact parameter.

\begin{figure}
\vspace{8.2cm}
\includegraphics{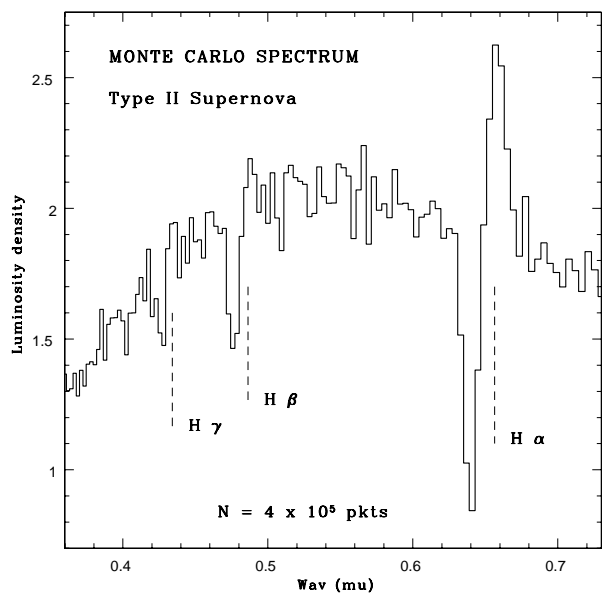}
\caption{Monte Carlo spectrum obtained at
iteration 20 in Monte Carlo calculation with ${\cal N} = 4 \times 10^{5}$
packets. The unit of luminosity density is $10^{45}$ erg s$^{-1}$ cm$^{-1}$.
The rest wavelengths of H$\alpha$, H$\beta$ and H$\gamma$ are indicated.}
\end{figure}

\section{Discussion and Conclusion}   

In Paper I, the radiative and collisional interactions of an atomic species
in statistical equilibrium with its environment was modelled in terms of
macro-atoms being activated and de-activated by the absorption and emission
of $e$-packets. In this paper, this macroscopic quantization 
has been extended to include interactions with the thermal pool
on the assumption of local thermal balance - i.e., ${\cal H} = {\cal C}$. 
(But note that departures from thermal equilibrium due to non-radiative
heating can be incorporated as described in Sect. 7.2 of Paper I.) 

	With this extension, the macro-atom formalism allows
MC transfer codes to obtain NLTE solutions for stratified atmospheres. As a
first test of this, this paper describes a MC code treating the formation
of the H spectrum in a Type II SN, with the emphasis being on demonstrating
the consistency of the technique as ${\cal N} \rightarrow \infty$, the
accuracy of the estimators of the various radiative rates for feasible  
values of ${\cal N}$, and the convergence behaviour of the constrained
$\Lambda$-iteration scheme. 

	Given that consistency has been demonstrated and that the 
geometry-independent $\Lambda$-iteration scheme converges, applications
to problems with arbitrary geometry and to multiple-species plasmas are
now in principle possible. But considering the detail and resolution required
to have impact on a current research topic, such problems will undoubtedly 
require a substantial increase in computational resources over those
deployed here
for a 1-D envelope of pure H. Even with the efficient estimators
of Sect. 6.2, each cell in the numerical grid must still be traversed by
numerous $r$-packets if adequate accuracy is to be achieved. This requirement
obviously mandates a huge increase in ${\cal N}$ for 2- and 3-D problems.
But also in 1-D problems for {\em static} atmospheres, narrow line widths 
will demand large values of ${\cal N}$ in order to get accurate b-b
transition rates.

	A further problem arises if the lower boundary of the
computational domain is at a large optical depth in the continuum. Each
$r$-packet's interaction history will then comprise numerous events and will
often end with a recrossing of the lower boundary, thus reducing the
number of $r$-packets traversing the cells in the outer atmosphere.

	This requirement for substantial computer resources suggests the
use of a computer with numerous parallel processors. Fortunately,
this MC technique is extremely well suited for such machines, especially those
with large shared memory. A MC transfer calculation can then be equally
divided between the processors, each of which carries its task to
completion with no exchanges of information or packets with other processors.
Following completion, the individual processors'
contributions to the estimator summations are added to obtain the
radiative rates for all cells, which are then equally divided among the
processors to solve the equations of statistical and thermal equilibrium.
These remarks suggest that close to the maximum theoretical efficiency
should be achievable.

	The long-term aim in devoloping the macro-atom formalism is
to obtain accurate NLTE solutions for multi-dimensional problems using MC
methods. But the technique is likely to be of more immediate use
in obtaining {\em approximate } NLTE solutions with precision sufficient
for current research problems. The reasons for optimism in this regard are
that, even without the converged stratifications of temperature and level
populations, the technique rigorously obeys radiative equilibrium and also
provides remarkably accurate emissivites (Paper I, Sect. 6). Thus, for some
problems, adequate accuracy will be achieved by using analytic ionization 
and excitation formulae, as in previous stellar wind and SNe codes, but then
computing the radiation field with this MC technique replacing the earlier
techniques that assumed coherent scattering (Abbott \& Lucy 1985) or downward
branching (Lucy 1999b). For yet higher accuracy, ionization could be
solved for while retaining an analytic excitation formula. Compared to the
full NLTE problem, this greatly reduces the demand for large ${\cal N}$ since
estimates of b-b rates are no longer required. Moreover, for each atomic
species, the statistical equilibrium
equations for perhaps thousands of levels are replaced by the ionization
equations for just $\sim 3-5$ ions.
Despite this enormous simplification, the anticipated loss of accuracy is
slight in view of the afore-mentioned accurate MC emissivities.

\section{Acknowledgement}

I am grateful to S.Sim for a discussion on the potential application of
parallel processors.

\end{document}